\documentclass{IEEEtran}
\usepackage{graphicx}
\usepackage{caption}

\usepackage{fixltx2e}
\usepackage{url}

\newcommand{\shorttitle}[1]%
{\markboth{Proceedings of the 1\MakeLowercase{$^{st}$} ICST, Greece 2009}{#1} }
\newcommand{\etal}{\MakeLowercase{\textit{et al. }}} 
                                                                                                              
\begin{document}
\title{X-ray experiments for Space applications in intermediate energy range}

\author{\IEEEauthorblockN{Vipin K. Yadav\IEEEauthorrefmark{1}\IEEEauthorrefmark{2},
Sandip K. Chakrabarti\IEEEauthorrefmark{3}\IEEEauthorrefmark{1}, 
Anuj Nandi\IEEEauthorrefmark{1}\IEEEauthorrefmark{2}, Sourav Palit\IEEEauthorrefmark{1}}.\\
\IEEEauthorblockA{\IEEEauthorrefmark{1} Indian Centre for Space Physics,
43 Chalantika, Garia Station Road, Kolkata - 700084, India. \\ vipin@csp.res.in, anuj@csp.res.in, sourav@csp.res.in}\\
\IEEEauthorblockA{\IEEEauthorrefmark{2} Indian Space Research
Organization HQ, New BEL Road, Bangalore - 560231, India.}\\
\IEEEauthorblockA{\IEEEauthorrefmark{3} S. N. Bose National Centre for
Basic Sciences, Block-JD, Sector-III, Salt Lake, Kolkata - 700098,
India. \\ Email: chakraba@bose.res.in}}

\shorttitle{Vipin \etal  X-ray Experiment}

\maketitle

\begin{abstract}

X-ray experiments in the intermediate energy range (1-50 keV) are
carried out at the Indian Centre for Space Physics (ICSP), Kolkata for
space application. The purpose is to carry out developmental studies of
space instruments to observe energetic phenomena from compact objects
(black hole and compact stars) and active stars and their testing and
evaluation.

\noindent
The testing/evaluation setup primarily consists of an X-ray generator,
various X-ray imaging masks, an X-ray imager and an X-ray spectrometer.
The X-ray generator (Mo target) operates in 1-50 kV anode voltage, and
1-30 mA beam current. A forty-five feet long shielded collimator is
used to collimate the beam which leads to the detector chamber. The
angular diameter of the X-ray beam at this distance becomes ≈ $\approx$
30 arc-sec. 

\noindent
Two types of imaging masks are being studied with: One is the
conventional Coded Aperture Masks (CAM) and other is the Fresnel
half-period zone-plates (ZPs) made of Tungsten. The latter has finer
zones of 40-50 microns rendering achromatic angular resolutions of
a few tens of arc-sec when two ZPs are kept at a distance of only 30 cm.
By increasing space between the ZPs one can achieve as high resolution
as necessary as long as the pointing accuracy is good enough. The Moire
fringe pattern produced by the composite shadows of two ZPs is inverse
Fourier transformed to obtain the X-ray source distribution. CAMs are
advantageous as they are single element devices, but the resolution
obtained is limited by their smallest pixel size. 

\noindent
A complementary metal-oxide semi-conductor (CMOS) detector connected to
a PC is used as the X-ray imager. This produces digitized image and can
be further analyzed. For spectroscopy a Si-PIN photo-diode based
detector is used. Several standard radioactive sources are used as
calibrators.

\noindent
Our setup has been extensively used in testing and evaluation of the
Roentgen Telescope (RT)-2 payloads which have been launched recently
(30th Jan. 2009). More experiments for improving imaging techniques are
being designed and tested.

\end{abstract}

\IEEEpeerreviewmaketitle

\section{Introduction}

\noindent
Study of compact objects are usually done by detecting high energy
photons emitted by accreting matter. Similarly stars in active period
also emit high energy radiation which may be detected and their
properties can be studied. Large number of telescopes have been launched
in the past to study both the compact objects, the sun and other stars.

\noindent
An energy range of importance belongs to the soft and intermediate
region (1-50 keV) where moderately active objects emit substantial
number of photons. Spectrophotometry and imaging in this range is our
priority. Though our laboratory is quite new, we have made our mark by
participating in building three payloads named RT-2 for the Russian
satellite KORONAS-FOTON. In this paper we present some of our studies.

\section{The X-ray Source}

\noindent
The X-ray beam is emitted from the generator having anode voltage in
the operating range of 0-50 kV and the beam current between 0-30 mA. The
X-ray target crystal is Molybdenum, which has good K$_{\alpha}$ and
K$_{\beta}$ line features. The X-ray beam when used at 45 feet away is
not quite a parallel beam but it can be treated as a quasi-parallel
(with little divergence) x-ray beam. The X-ray beam diameter ($l$) is 2
mm and the X-ray collimator length ($L$) is 45 feet (13716 mm). Hence,
the beam divergence at a distance of 45 feet is $\sim 30$ arcsec. The
X-ray beam from the source to a distance of 45 feet is guided through a
7.5 cm diameter Aluminium pipe-line. This pipe-line is covered
throughout its length by a 2 mm thick lead sheet for shielding. 

\section{The Imaging Devices - Zone Plates}

\begin{figure}[h]
\includegraphics[height=2in,width=3.5in]{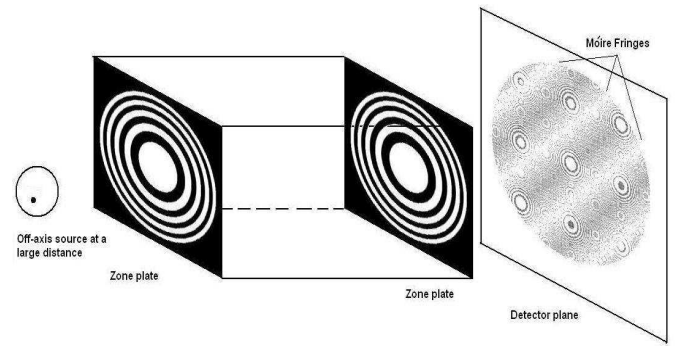}
\caption{The zone-plate working principle.}
\label{zpp}
\end{figure}

\noindent
ICSP has been toying with both the conventional Coded Aperture Masks
(CAMs) as well as Fresnel Zone Plates (FZPs) \cite{young,mertz}. CAMs
are widely used, but higher resolutions can be achieved by FZPs as well.
Two aligned zone plates will cast Moire fringes when the source is
off-axis (Figure \ref{zpp}). The fringe pattern can be appropriately
inverse Fourier transformed to get back the source pattern.

\begin{figure}[h]
\includegraphics[height=2.0in,width=1.6in]{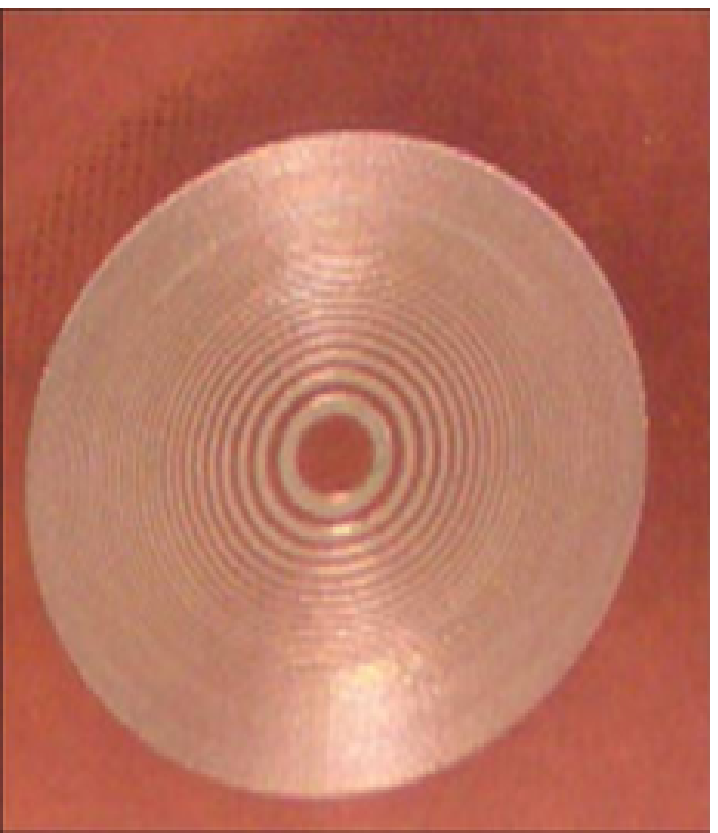} \hspace{5mm}
\includegraphics[height=2.0in,width=1.6in]{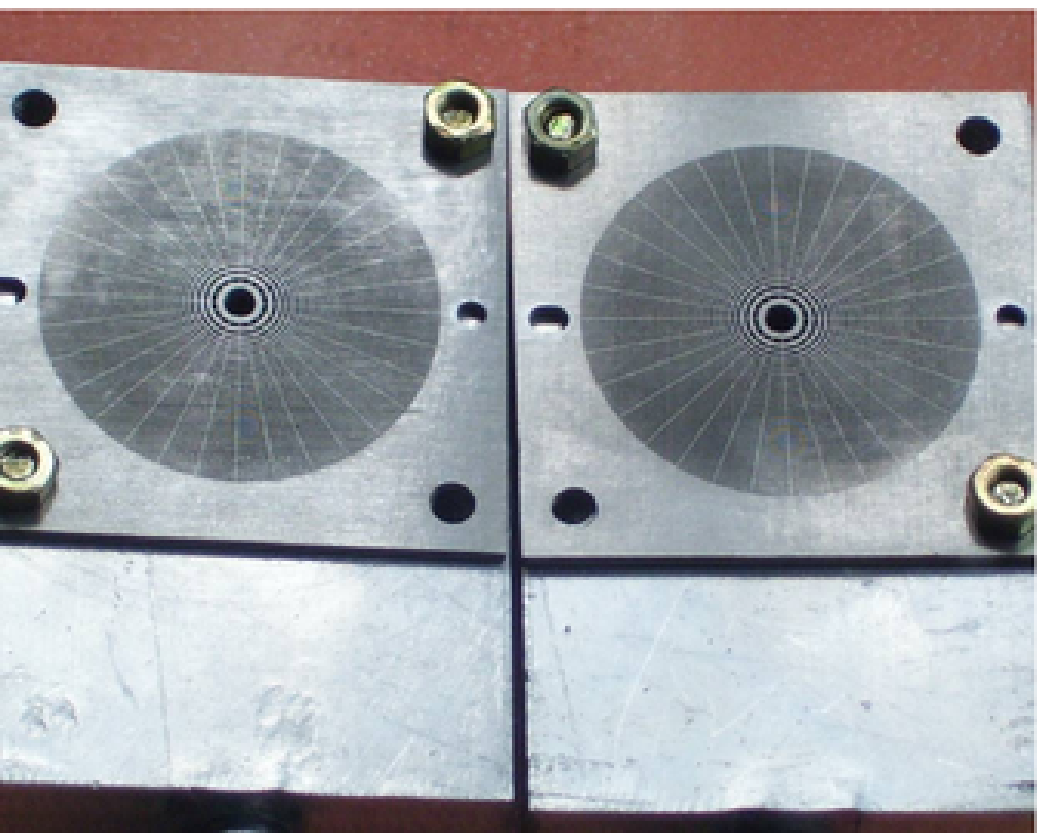}
\caption{The Fresnel zone-plates made of copper (left) and tungsten
(right).}
\label{cuwzp}
\end{figure}

\noindent
At ICSP, we have designed zone plates and got them fabricated using
copper and tungsten. The copper zone plates can be used only for soft
X-rays but the latter can be used for hard X-rays also. The advantage
with copper zone plates is that they are less expensive to build and
can be made in large sizes also. In figure \ref{cuwzp}, the various
types of these zone plates are depicted.

\section{The Detection System}

\noindent
The detector systems used to record the experimental results are X-ray
films, Amptek made Si-PIN detector with 25 $\mu$m thick Be window for
low energy X-rays (Energy resolution at 5.9 keV $Fe_{55}$, 145 eV FWHM),
Rad-icon made RemoteRadEye CMOS imager (1024 pixels $\times$ 1024
pixels) and Orbotech made CZT (Cadmium Zinc Telluride) detectors.

\begin{figure}[h]
\includegraphics[height=2.5in,width=3.40in]{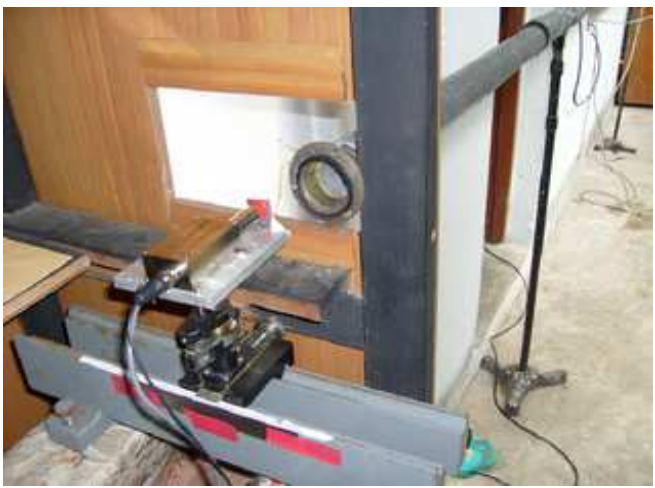}
\caption{The Si-PIN detector installed at the end of 45 feet shielded
x-ray beam-line.}
\label{sipin}
\end{figure}

\noindent
In Figure \ref{sipin} we have shown the placement of the Si-PIN
photo-diode detector at the end of 45 feet long X-ray beam-line during
the experiments conducted.

\noindent
As can be seen from the figure, the Si-PIN detector is mounted on a
custom built optical bench and the detector can be linearly shifted by
1 mm if required. 

\begin{figure}[h]
\includegraphics[height=2.5in,width=3.40in]{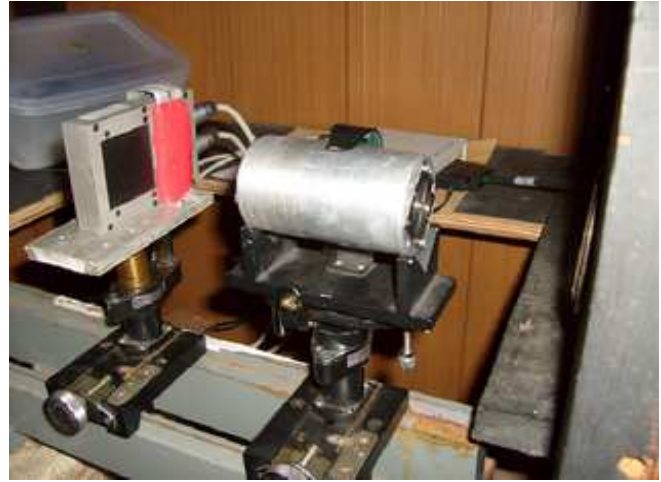}
\caption{The CMOS detector installed at the end of 45 feet shielded
X-ray beam-line.}
\label{cmos}
\end{figure}

\noindent
In Figure \ref{cmos} we have shown the placement of the CMOS detector
at the end of 45 feet long X-ray beam-line during the testing campaign
\cite{skc}.

\noindent
Some of the experimental results obtained using these detecting
devices are presented the following sections.

\section{Experimental Results}

\begin{figure}[h]
\includegraphics[height=2.0in,width=1.6in]{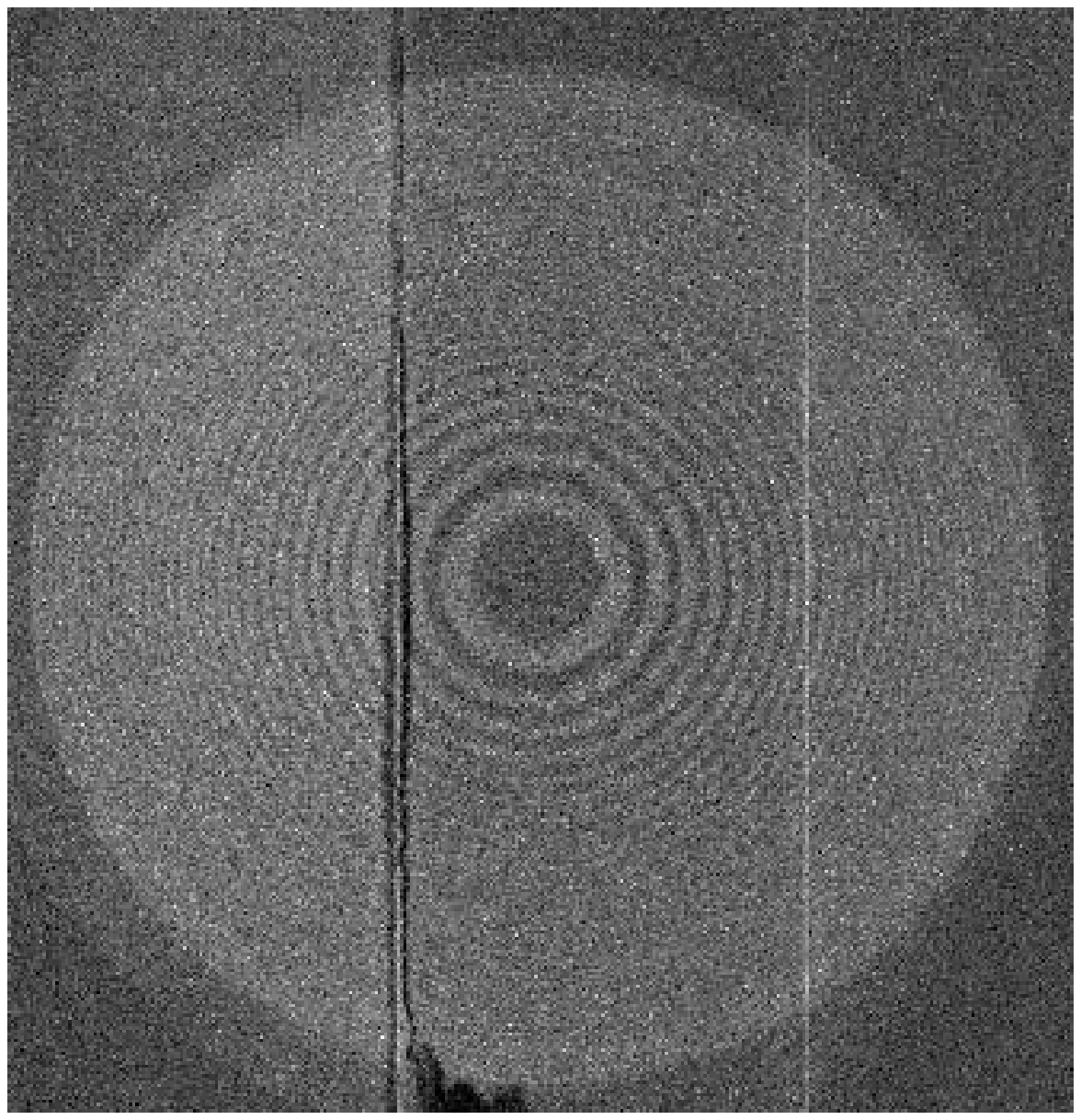} \hspace{5mm}
\includegraphics[height=2.0in,width=1.6in]{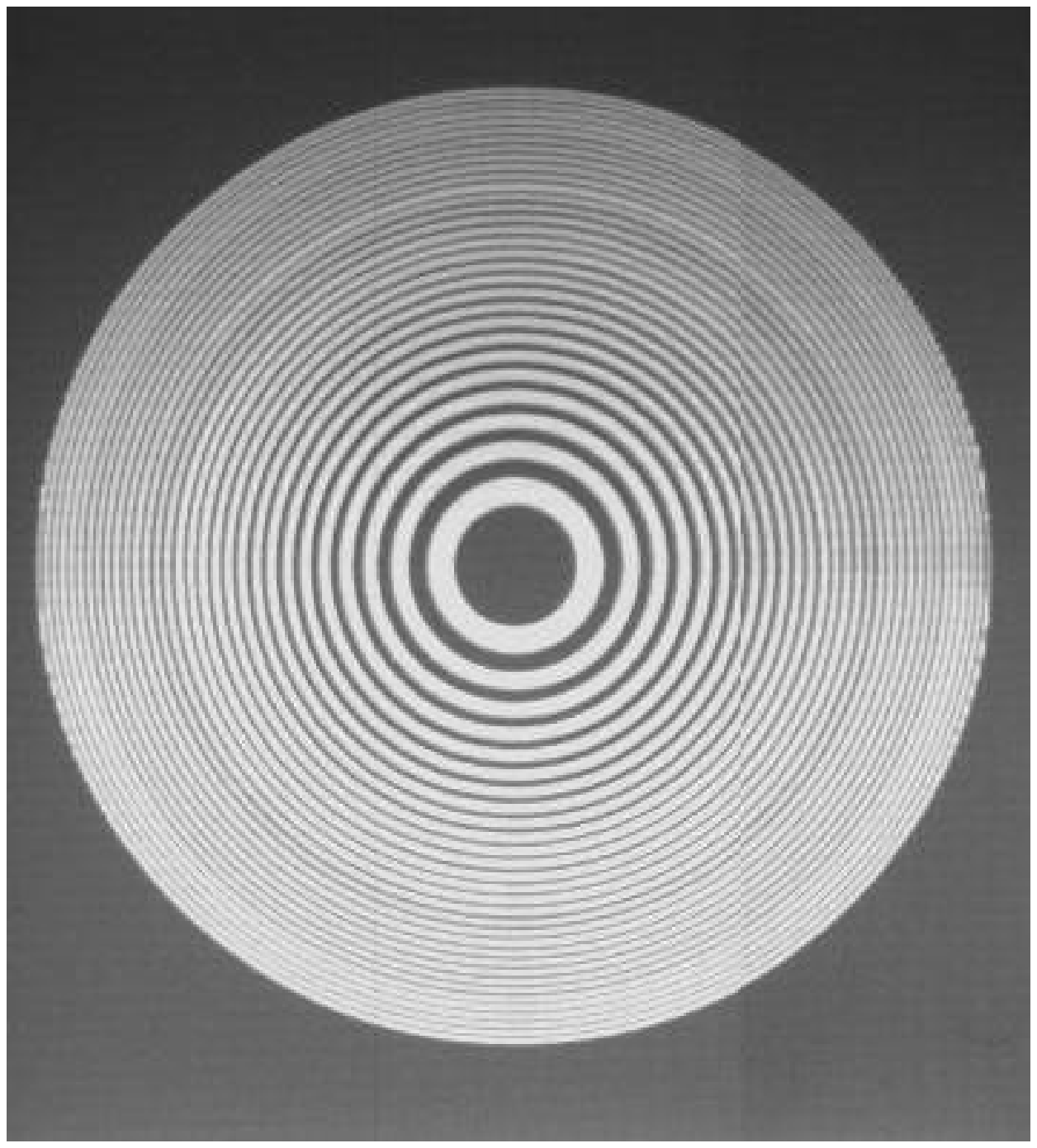}
\caption{The CMOS detector images of the copper (70 $\mu m$ thick) zone
plate (left). The CMOS detector images of the copper zone plate coated
with 100 $\mu m$ gold (right).}
\label{cuzp}
\end{figure}

\noindent
In the left side of the Figure \ref{cuzp}, the X-ray image of the copper
Zone plate is shown as obtained at a distance of 45 feet by the CMOS
detector. As expected the image is blur due to poor blocking of X-rays
by the copper. To increase the X-ray absorption by the copper zone
plates, gold coating is done on to them. The gold plated zone plate
clearly is sharper and is blocking hard X-rays as well. This is shown in
the right hand side figure \ref{cuzp}.

\begin{figure}[h]
\includegraphics[height=2.0in,width=1.6in]{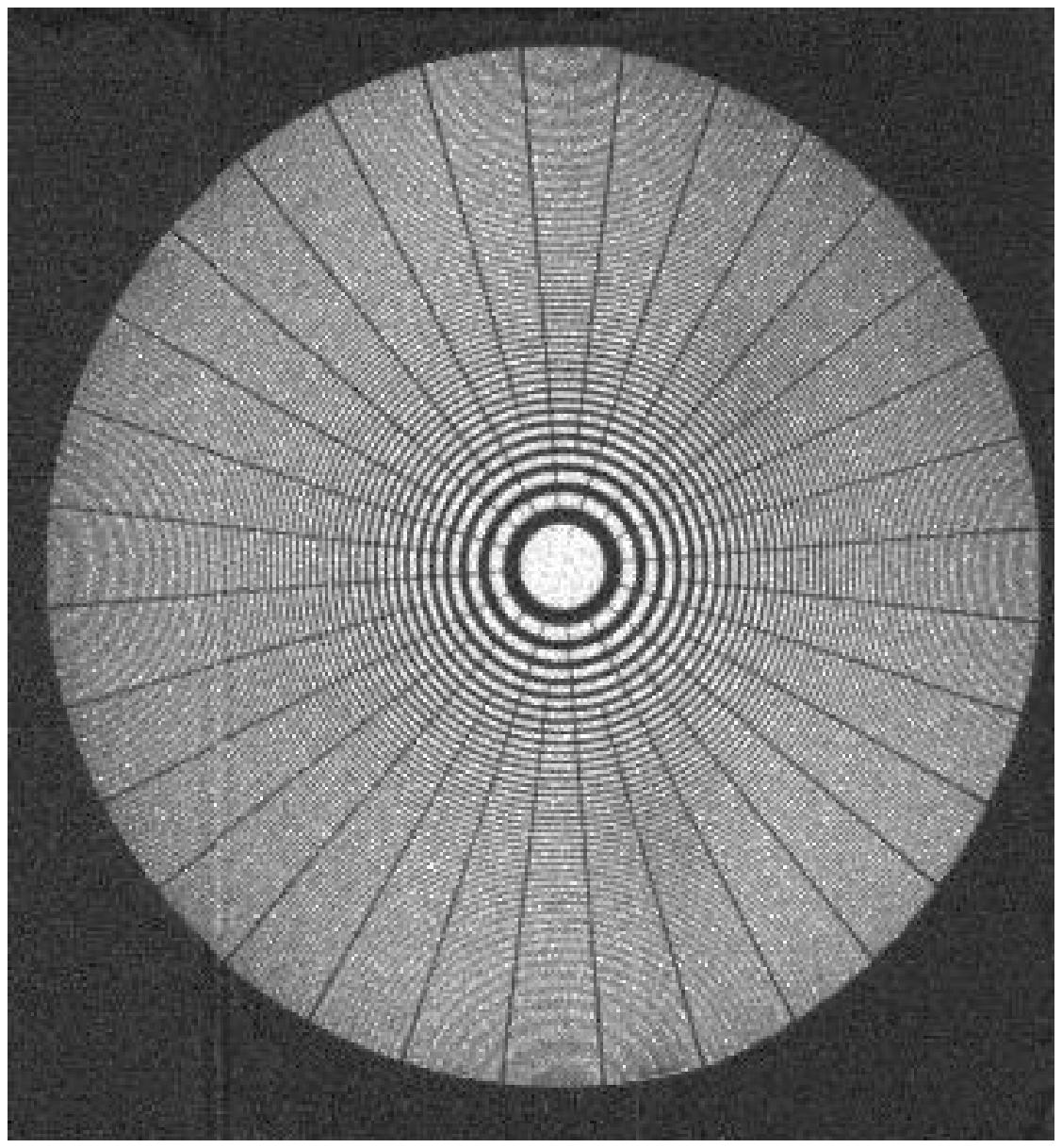} \vspace{5mm}
\includegraphics[height=2.0in,width=1.6in]{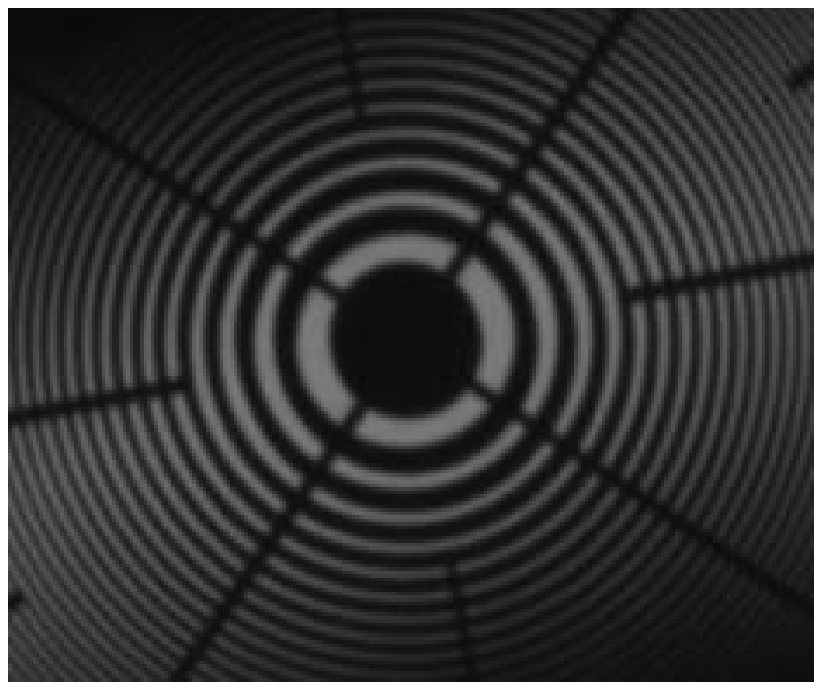}
\caption{An image of positive zone plate (left) and a close-up of a
negative zone plate (right). The images are taken by the CMOS detector.}
\label{posnegzp}
\end{figure}

\noindent
Experiments are performed with all possible combinations of tungsten
zone plates available which include positive (central zone transparent)
as well as negative (central zone opaque). The left side photo in
figure \ref{posnegzp} shows one such positive zone plate. In the same 
figure on the right side the zoomed in region of a negative zone plate.
The output in both these photos is taken on our CMOS detector.

\begin{figure}[h]
\includegraphics[height=3.4in,width=3.40in]{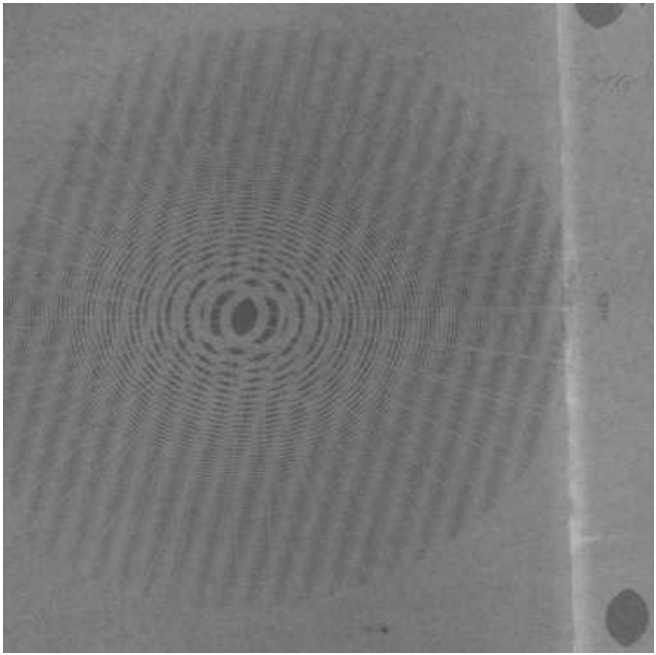}
\caption{Moire fringes obtained from two tungsten zone plates on a
photographic plate.}
\label{pp-moire}
\end{figure}

\noindent
Figure \ref{pp-moire} shows the Moire fringes obtained on photographic
plates by two zone plated separated by 10 cm and the X-ray source is
off-axis.

\noindent
It is to be noted here that the photographic plates provides very good
resolution. 

\begin{center}
\begin{figure}[h]
\includegraphics[height=3.3in,width=3.3in]{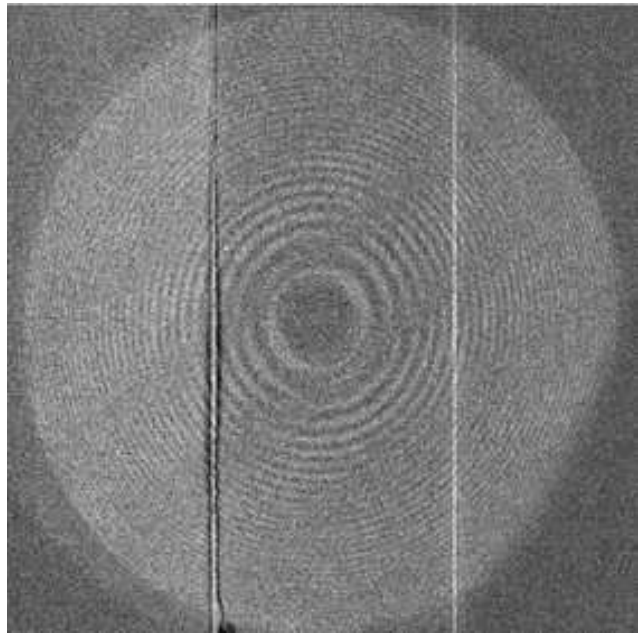}
\caption{Moire fringes obtained with two copper zone plates placed back
to back.}
\label{cuzp-bk}
\end{figure}
\end{center}

\noindent
The moire fringes as also obtained by using two zone plates made from
copper and the CMOS detector. However, in this experiment, the zone
plate pair is kept in contact with each other. The CMOS image output is
faint and is shown in figure \ref{cuzp-bk}. The latter fringes are very
faint. Therefore, it is decided that in future, the gold plated copper
zone plates (Figure \ref{cuzp}) are to be used to obtain sharper
fringes.

\begin{figure}[h]
\includegraphics[height=3.3in,width=3.3in]{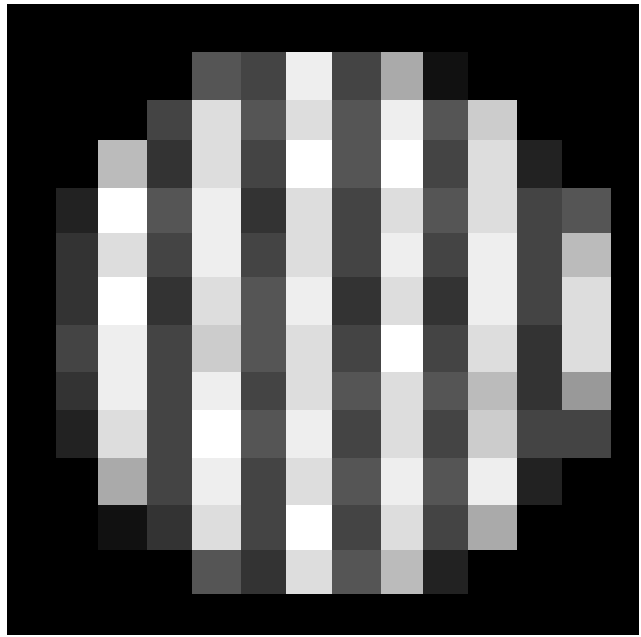}
\caption{Moire Fringes of the X-ray source obtained on the CZT
detector.}
\label{czt}
\end{figure}

\noindent
In Figure \ref{czt}, we show the Moire fringes on CZT detectors
(having a pixel size of 0.25 cm) on the left and CMOS detector (having
a pixel size of 50 $\mu m$) on the right. We use special IDL programme
to obtain source details from these fringes \cite{sourav}. 

\begin{figure}[h]
\includegraphics[height=3.3in,width=3.3in]{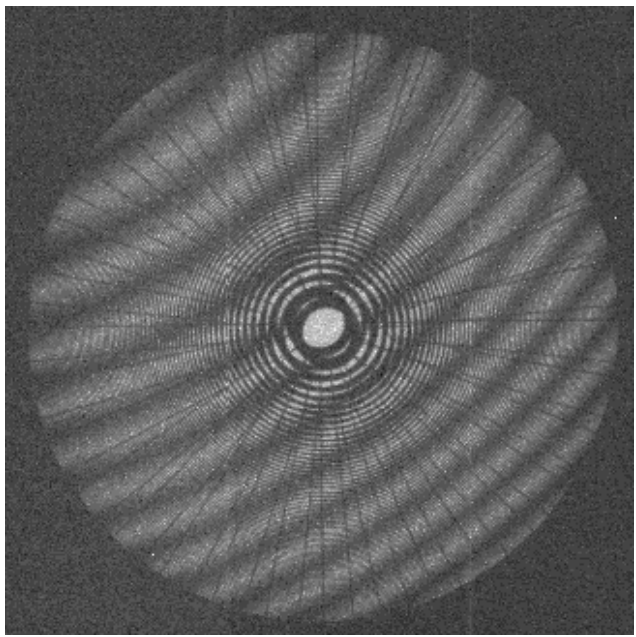}
\caption{Moire Fringes for the X-ray source obtained on the CMOS
detector.}
\label{cmos-moire}
\end{figure}

\noindent
In figure \ref{cmos-moire}, the same source image taken with CMOS
detector is shown.

\section{Major test and evaluation at ICSP: RT-2 payloads}

\begin{figure}[h]
\includegraphics[height=2.75in,width=3.3in]{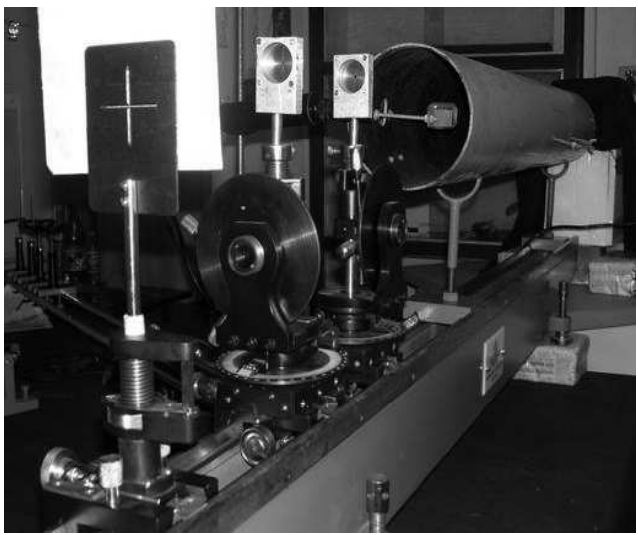}
\caption{The Tungsten zone plate pair alignment set-up}
\label{wzpalign}
\end{figure}

\noindent
In Figure \ref{wzpalign}, we show our precision optical bench set up
which is used to align components which require alignments accurate to
arcseconds using steady laser beams.

\begin{figure}[h]
\includegraphics[height=2.8in,width=3.3in]{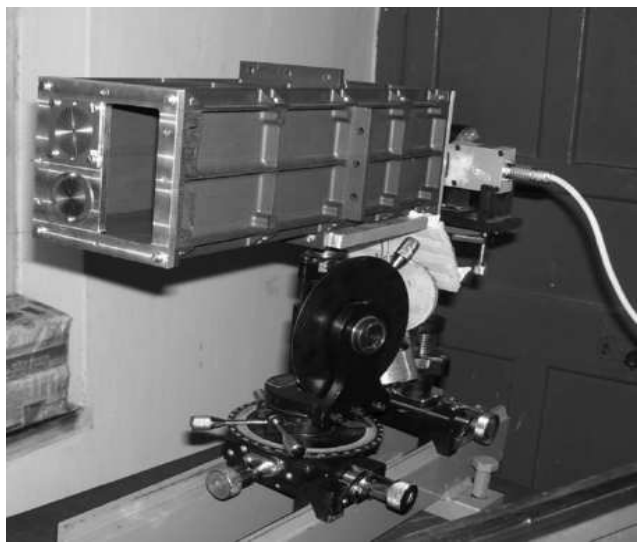}
\caption{The RT-2 collimator installed on the x-ray system test bench
during testing.}
\label{rt2}
\end{figure}

\noindent
This setup has been used to test the collimators with zone plates for
the RT-2/CZT payload which has been sent aboard Russian satellite
KORONAS-FOTON. A typical testing configuration of the collimator is
shown in figure \ref{rt2}. Two different sets of zone plates mounted on
the collimator can be seen in this figure.

\noindent
We have already tested and evaluated another payload namely RT-2/S (a
phoswich detector) which is going to work in the x-ray energy range of
15 - 150 keV. The instruments (RT-2/S and RT-2/CZT) onboard
KORONAS-FOTON satellite are launched on January 30, 2009.

\section{Other activities}

\noindent
Other activities of the ICSP X-ray laboratory includes developmental
work on Si-PIN photo-diode based X-ray detector, drift detectors, X-ray
counters etc. These are being tested in balloons by ICSP scientists.

\section{Concluding remarks}

\noindent
ICSP X-ray Laboratory is equipped with instruments to develop and test
and evaluate equipments for space application. We concentrate only on
the soft and intermediate X-ray range i.e., 1 - 50 keV. We have
participated in RT-2 payloads for the Russian KORONAS-FOTON satellite.
We are in a position to propose future missions for continuous
spectrophotometry of black holes.

\section*{Acknowledgment}

The authors would like to thank ISRO for providing finance towards the
experiments. The research of S. Palit is supported by CSIR.

\end{document}